\documentclass{article}

     \PassOptionsToPackage{numbers, compress}{natbib}

\usepackage[final]{neurips_2023_ml4ps}

\usepackage[utf8]{inputenc} 
\usepackage[T1]{fontenc}    
\usepackage{hyperref}       
\usepackage{url}            
\usepackage{booktabs}       
\usepackage{amsfonts}       
\usepackage{nicefrac}       
\usepackage{microtype}      
\usepackage{xcolor}         

\usepackage{graphicx, float, amsmath, array} 

\newcommand{\bm}[1]{\boldsymbol{#1}}
\setcitestyle{square,numbers}

\title{Accelerating Flow Simulations using Online Dynamic Mode Decomposition}

\author{%
  Seung Won Suh \\
  Department of Mechanical Science and Engineering\\
  University of Illinois, Urbana-Champaign\\
  Champaign, IL \\
  \texttt{suh29@illinois.edu} \\
  \And
  Seung Whan Chung \\
  Lawrence Livermore National Laboratory \\
  Livermore, CA \\
  \texttt{chung28@llnl.gov} \\
  \AND
  Peer-Timo Bremer \\
  Lawrence Livermore National Laboratory \\
  Livermore, CA \\
  \texttt{bremer5@llnl.gov} \\
  \And
  Youngsoo Choi \\
  Lawrence Livermore National Laboratory \\
  Livermore, CA \\
  \texttt{choi15@llnl.gov}
}

\begin{document}

\maketitle

\begin{abstract}
  We develop an on-the-fly reduced-order model (ROM) integrated with a flow simulation, gradually replacing a corresponding full-order model (FOM) of a physics solver. Unlike offline methods requiring a separate FOM--only simulation prior to model reduction, our approach constructs a ROM dynamically during the simulation, replacing the FOM when deemed credible. Dynamic mode decomposition (DMD) is employed for online ROM construction, with a single snapshot vector used for rank-1 updates in each iteration. Demonstrated on a flow over a cylinder with $\mathrm{Re}=100$, our hybrid FOM/ROM simulation is verified in terms of the Strouhal number, resulting in a 4.4 times speedup compared to the FOM solver.
\end{abstract}

\section{Introduction}

Iterative solvers for systems with a large number of degrees of freedom (DOFs) can be found in almost all areas of computational science and engineering. Physics simulations based on partial differential equations (PDEs) are one of the classic examples, where the solution is represented with a finite number of DOFs in space and marches in time as matrix equations are solved over multiple time steps. Optimization is another example where one first solves a primal problem to evaluate a quantity of interest and then solves its dual to compute sensitivities iteratively for gradient descents, e.g., PDE-constrained optimization or training of neural networks.

Although FOM--based solvers may generate predictions of high fidelity, they easily get computationally infeasible if one tries to resolve a much wider range of scales. At times when it is more important to get a rough estimate rapidly, one might need to consider building ROMs which are less accurate but cheaper than FOMs. Often used for building a ROM are modal analysis techniques such as the DMD, which is a data-driven algorithm that extracts the dynamics of the temporal evolution of a system \cite{schmid2010dynamic}.

Here we present an efficient workflow for constructing the ROM for flow simulations. Before getting started, we emphasize that such traditional reduced-order modeling is not different from the techniques used to reduce dimensionalities in the field of machine learning. For instance, the principal component analysis (PCA) is merely another name of the proper orthogonal decomposition (POD) \cite{berkooz1993proper}, which is widely used for reduced-order modeling in physics simulations \cite{carlberg2018conservative,choi2020gradient,choi2019space,choi2021space,hoang2021domain,mcbane2021component,choi2020sns,copeland2022reduced,kim2021efficient,fries2022lasdi,lauzon2022s,cheung2023local,mcbane2022stress,tsai2023accelerating}. Therefore, our work of reducing the size of flow simulations using the DMD can be regarded adjacent to machine learning--based surrogate modeling and the workflow that we are presenting is not limited to the DMD but can be applied to any type of technique for dimensionality reduction.


A plethora of works have already shown the capability of the DMD to learn the nonlinear dynamics including that of fluid flows \cite{kutz2016dynamic}, pore-collapse \cite{cheung2023datascarce}, and radiative diffusion problem \cite{huhn2023parametric}. However, most of the studies are limited to offline training. That is, many full-order model simulations need to be run to generate the whole set of samples, and then the ROM is trained separately from the simulation using those samples. This may be undesirable when the dimensionality of the underlying parameter space is high, leading to the curse of dimensionality and requiring a large number of sampled training points. In such scenarios, efficiency can be enhanced by adopting an on-the-fly approach—interrupting the full-order model simulation and substituting the reduced-order model (ROM) at the relevant time step—rather than adhering to separate offline training and online inference steps. Via such a hybrid FOM/ROM method, we anticipate achieving two things at once: (1) accelerating the simulation itself, and (2) constructing a ROM that can be reused later for other simulations having similar setups, e.g., those for parametric studies. The existing on-the-fly data-driven approach includes projection-based reduced order modeling on PDE-constrained optimization \cite{choi2019accelerating, wen2023globally}. Similar approaches of using online POD have also been introduced by \cite{rocha2020adaptive,huang2023predictive}, as well as those using artificial neural networks by \cite{fritzen2019fly} and \cite{rocha2021fly}.

In this work, we demonstrate a workflow for hybrid FOM/DMD flow simulations---which has not yet been attempted up to the authors' knowledge---where the FOM is gradually replaced by the DMD trained on the fly. Each iteration consists of two key steps: (1) to update the DMD and (2) to replace the FOM with the DMD if possible. In addition, if the DMD is thought to have been converged, we end the simulation and directly make a prediction at the final time step, skipping all the remaining time steps and thereby significantly reducing the cost. This hybrid method has been applied to the simulation of a flow over a cylinder, followed by a comparison to a FOM--only simulation.

\section{Methodology}

\subsection{Workflow}

\begin{figure}
    \centering    \includegraphics[width=\linewidth]{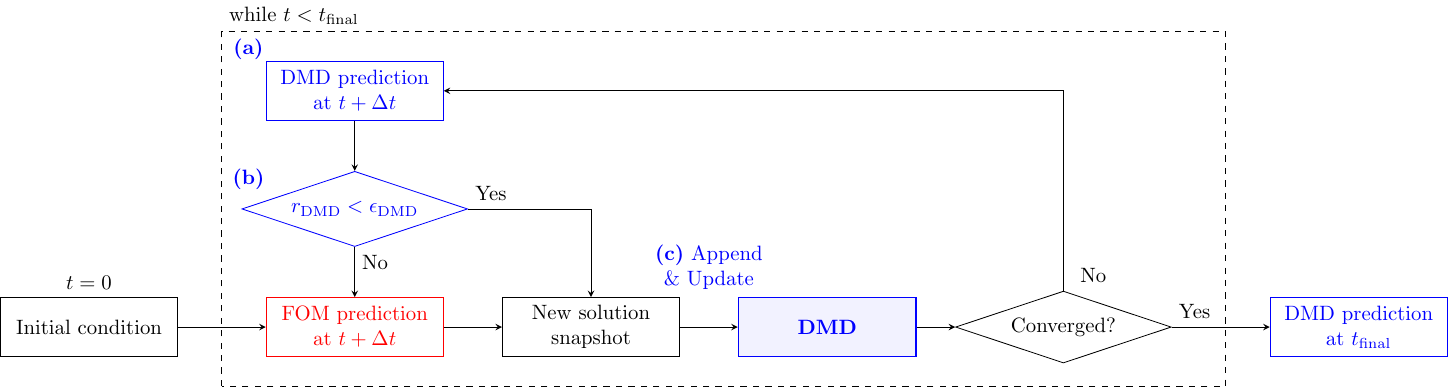}
    \caption{Workflow of the hybrid FOM/DMD simulation of a dynamical system.}
    \label{workflow}
\end{figure}

Figure \ref{workflow} describes the workflow of a hybrid FOM/DMD simulation. Like standard solvers for time-dependent systems, we start the time integration with an initial condition. At the beginning of the first time step, there is no DMD initialized so we call the FOM and solve the system. Once we get a new snapshot of the solution, we use it to initialize the DMD, and then proceed to the next time step. At the beginning of the second time step, we first make a prediction on the solution at the second time step with our initialized DMD. If it is accurate enough, i.e. the residual of the prediction is smaller than the tolerance, we use that prediction as our new snapshot and move on to the next time step without calling the FOM. If not, we call the FOM and use the FOM prediction as the new snapshot to update the previously-trained DMD model in a rank-1 update fashion. While the DMD gradually replaces the FOM, we also check whether the DMD itself has converged and if so, we exit the loop and go directly to $t_\mathrm{final}$ to make a prediction.

\subsection{Error Estimator}

One way to evaluate the accuracy of the prediction at the $k$-th time step made by the constructed DMD is to measure the local truncation error $\bm e_\mathrm{DMD}^k = \bm u_\mathrm{DMD}^k - \bm u_\mathrm{FOM}^k$ where $\bm u^k$ is the prediction of the solution at the $k$-th time step by either the DMD or the FOM. However, computing the error should be avoided since it requires solving the FOM for $\bm u_\mathrm{FOM}^k$. Instead, we compute the residual as a representative of the local error. If a FOM solves a linear system $\bm A \bm u = \bm b$, the residual is defined as $\bm r(\bm u) = \bm b - \bm A \bm u$ which does not require solving the matrix equation. As $\bm r(\bm u_\mathrm{DMD}) = -\bm A \bm e_\mathrm{DMD}$, the residual is a linear map of the error and is a valid estimator of the error. Likewise, for nonlinear systems, one can derive residual-based error bounds, indicating that the residual norm is a good error indicator. In a nutshell, whenever we make a prediction with the DMD, we compute the residual as $\bm r (\bm u_\mathrm{DMD})$ and if its norm is smaller than a certain level of tolerance $\epsilon_\mathrm{DMD}$, we trust the DMD and use its prediction as our new snapshot. Note that the residual depends on the numerical scheme used for the FOM.

\subsection{Online DMD}
The DMD approximates either linear or nonlinear dynamics using a linear operator. Consider a dynamical system that evolves over time as $\bm u^{k+1} = \bm F(\bm u^k)$ where $\bm u^k \in \mathbb R^N$ is an $N$-dimensional state vector at time steps $k = 0, 1, \cdots, n$ and $\bm F: \mathbb R^N \to \mathbb R^N$ is an operator that describes the temporal evolution of the state. Then we can construct two matrices composed of snapshot vectors $\bm U^+ = \left[ \bm u^n \vert \bm u^{n-1} \vert \cdots \vert \bm u^1 \right] \in \mathbb R^{N \times n}$ and $\bm U^- = \left[ \bm u^{n-1} \vert \bm u^{n-2} \vert \cdots \vert \bm u^0 \right] \in \mathbb R^{N \times n}.$ The DMD matrix is then given as $\bm A = \bm U^+ (\bm U^-)^\dagger \approx \bm U^+ \bm R \bm S^{-1} \bm L^* = \bm A_r \in \mathbb R^{N \times N}$ where $\bm U^- = \bm L \bm S \bm R^*$ is the truncated SVD at rank $r \leq n$ ($\bm L \in \mathbb R^{N \times r}, \bm S \in \mathbb R^{r \times r}, \bm R \in \mathbb R^{m \times r}$) and ${}^\dagger$ is the Moore-Penrose inverse of a matrix. Furthermore, the full-rank reduced DMD matrix is defined as $\tilde{\bm A}_r = \bm L^* \bm U^+ \bm R \bm S^{-1} \in \mathbb R^{r \times r}$ and the projected DMD modes are $\bm \Phi = \bm L \bm V \in \mathbb R^{N \times r}$ where $\tilde{\bm A}_r = \bm V \bm \Lambda \bm V^{-1}$ is the eigenvalue-decomposition of $\tilde{\bm A}_r$. Once the DMD matrices are constructed, one can predict the solution at any time $t$ based on the analytical form $\bm u(t) = \bm \Phi \bm \Lambda^{t/\Delta t} \bm \Phi^\dagger \bm u(0).$ Hence, the DMD forms a ROM that makes a prediction at a constant time complexity of $\mathcal O(Nr)$, independent of $t$.

In our hybrid approach, updating the DMD matrices by a single rank is sufficient as only a single snapshot is appended to the snapshot matrix every iteration. Incremental SVD by \cite{brand2006fast} is an algorithm that efficiently makes a rank-1 update to the SVD of a matrix. The essence of the algorithm is to avoid costly matrix-matrix multiplications and utilize matrix-vector multiplications to get a speedup by $\mathcal O(r)$. We can employ the ideas from incremental SVD to efficiently make rank-1 updates on the DMD matrices. Although several authors have already introduced the idea of online DMD \cite{matsumoto2017fly, zhang2019online}, ours differ from theirs as \cite{zhang2019online} is based on the Sherman--Morrison formula and \cite{matsumoto2017fly} used the modified version of the original algorithm \cite{brand2006fast}, while we stick to the original version to update the DMD matrices. In the end, each DMD update costs $\mathcal O(Nr)$, which is more efficient than naively constructing the matrices from scratch that costs $\mathcal O(Nnr)$ for $N \gg n \gg r$. Details of the implementation are not presented in this paper due to the limited space.

\section{Numerical Experiments \& Results}

The prescribed hybrid workflow was applied to simulate a three-dimensional incompressible flow over a cylinder. The cylinder has a diameter of $D$ with its axis along the $z$-direction at $(x_c, y_c) = (5D, 0)$. The computational domain was defined as $\Omega = [0, L_x] \times [-L_y/2, L_y/2] \times [-L_z/2, L_z/2]$ with the dimensions $(L_x, L_y, L_z) = (25D, 4.1D, 4.1D)$. Dirichlet boundary conditions were set for the velocities at the inflow ($x=0$) and the top and bottom walls ($y=\pm L_y/2$) as $(u,v)=(U,0)$, and pure Neumann boundary conditions were set elsewhere. For the pressure, pure Neumann boundary conditions were applied at the inflow and the top and bottom walls, and the Dirichlet boundary condition was used at the outflow $(x=L_x)$ as $p=0$. The kinematic viscosity $\nu$ was set to match the Reynolds number $\mathrm{Re} = UD/\nu = 100$ where we can anticipate mild vortex streets. Although the top and bottom boundaries were quite close to the cylinder, we were able to observe a distinct vortex-shedding structure throughout the simulation.

An incompressible flow solver in MFEM \cite{mfem-web} was used as the FOM, which uses a nodal $H^1$ finite element discretization with polynomial order of $3$ in space and an implicit (of order 3) / explicit (of order 3) time integration of the viscous and convective terms in the momentum equation, respectively. More detailed explanations of the numerical scheme can be found in \cite{mfem-web, tomboulides1997numerical, franco2020high}.

An unstructured coarse baseline mesh was further refined within the solver where the total number of DOFs was 585,732. A preliminary high-fidelity FOM simulation was first run for $t^*=Ut/D=50$ starting from the zero flow until it was thought to have entered the limit cycle of the vortex shedding. The time step size was adjusted so the maximum CFL number remained around 0.2 throughout the simulation. Then, the final snapshot of the solution was used to initialize flow fields for the hybrid and restarted FOM--only simulations, which were both run until $t^*=100$. For a fair comparison, the level of fidelity of the restarted FOM--only simulation has been lowered down to that of the DMD: the tolerance level of the conjugate-gradient solver ($\epsilon_\mathrm{CG}$) for the FOM--only simulation and the tolerance level for the accuracy of the DMD ($\epsilon_\mathrm{DMD}$) were both set to $10^{-5}$. Note that the FOM for the hybrid simulation retains the high fidelity of $\epsilon_\mathrm{CG}=10^{-12}$.


\begin{figure}
    \centering
    \begin{tabular}{m{0.6\textwidth} m{0.3\textwidth}}
        \includegraphics[width=\linewidth]{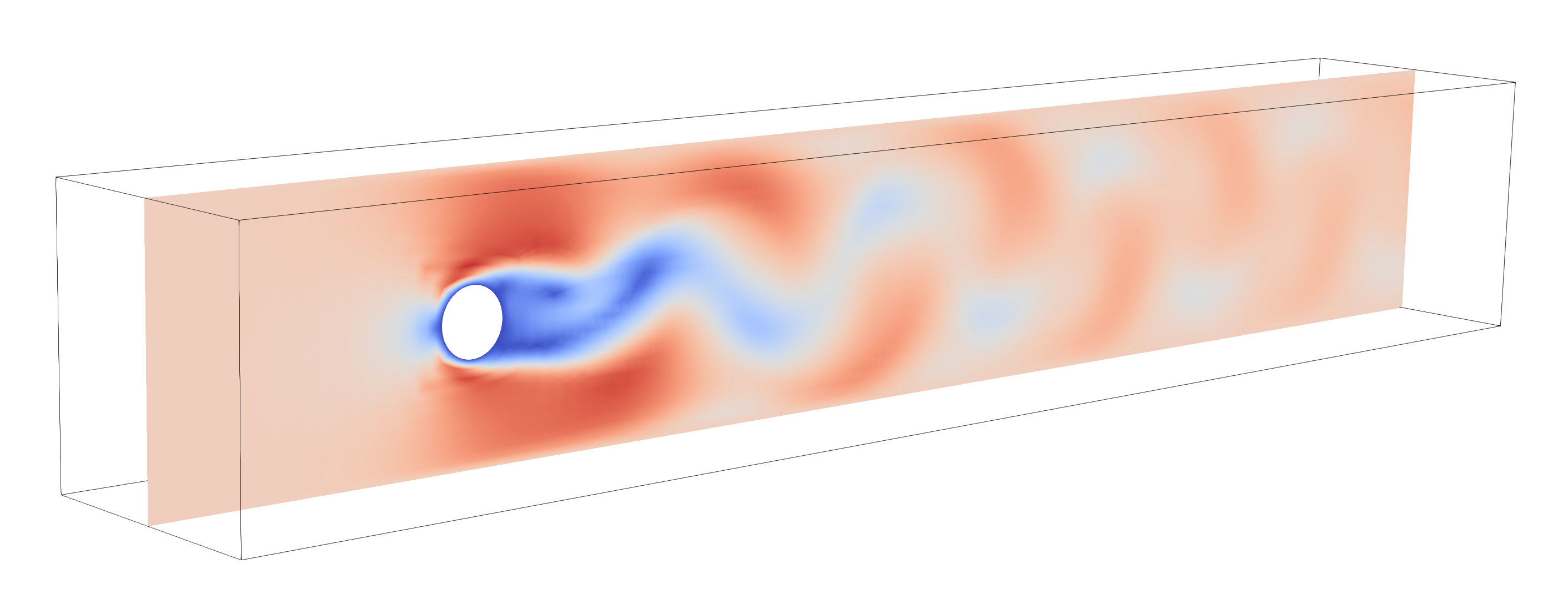} &
        \includegraphics[width=\linewidth]{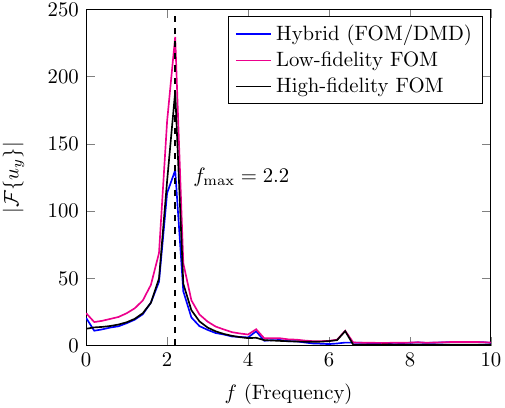}
    \end{tabular}
    \caption{Velocity magnitude plotted on a 2-D slice normal to the $z$-axis at $t^*=50$ (left); temporal spectrum of $u_y$ sampled downstream.}
\end{figure}

Temporal spectra of the $y$-velocity at a single point $(x,y,z)=(10D,0,0)$ from the hybrid and a high-fidelity ($\epsilon_\mathrm{CG}=10^{-12}$) FOM--only simulations were compared for verification. Frequencies of the vortex shedding predicted by the two simulations were identical within the precision of the discrete values of frequency. The Strouhal number $\mathrm{St} = fD/U$---the dimensionless frequency of the vortex shedding---was 0.22 in both simulations, which is around the reported value of 0.16 at $\mathrm{Re}=100$ \cite{williamson1989oblique}. The discrepancy is thought to be due to the boundary effect.

\begin{figure}
    \centering
    \includegraphics[width=0.8\linewidth]{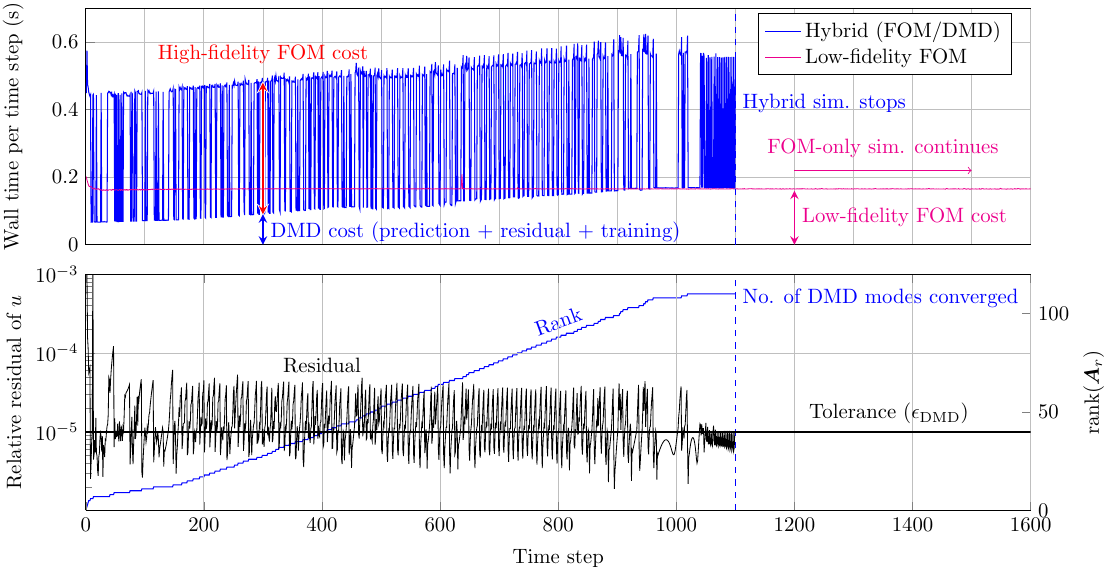}
    \caption{Wall time per time step: hybrid vs. FOM (top); history of the relative residual of $u_\mathrm{DMD}$ and the rank of $\bm A_r$ (bottom). The horizontal axis only shows until the 1,600$^\text{th}$ time step out of 10,000 time steps of the FOM--only simulation.} \label{plot}
\end{figure}

Figure \ref{plot} shows the cost of the two simulations and the history of the residual of $u$ from the hybrid simulation. Notice that at the early stage of the simulation, the residual keeps going up and down around the level of tolerance ($10^{-5}$) so the FOM is frequently called. As the simulation goes on, the model gets more accurate and hits below the level of tolerance more often, leading to less frequent calls of the FOM. This depicts how the FOM is gradually replaced by the DMD throughout the simulation as we expected. In total, the hybrid and the low-fidelity FOM--only simulations took 375s and 1648s for $10^4$ time steps, respectively, thereby showing about a 4.4-times speedup. Moreover, even when we choose not to skip all the remaining time steps of the simulation, a single time-stepping using the converged DMD is 5.77 times faster than using the low-fidelity FOM so the total cost of our approach will remain cheaper.

\begin{table}
    \centering
    \begin{tabular}{c c c c}
        & Hybrid (FOM/DMD) & Low-fidelity FOM & \textbf{Speedup} \\
        \hline
        Total simulation time (s) & 375 & 1648 & \textbf{4.39} \\
        Single-$\Delta t$ prediction time (s) & 0.0286 & 0.165 & \textbf{5.77}
    \end{tabular}
    \bigskip
    \caption{Two different types of speedups, in terms of the total simulation time and the cost of a single time-stepping. The total simulation time for the FOM/DMD approach includes the time for making a prediction, checking accuracy, and training. The cost of a single time-stepping only measures the time spent on making a prediction, based on the converged DMD.}
\end{table}

\section{Conclusion}
We used the DMD to reduce the dimension of the FOM that is evaluated iteratively during flow simulations, where the DMD is constructed on the fly and gradually replaces the FOM. This has been demonstrated on a simulation of a flow over a cylinder, and we were able to gain a 4.4-times speedup.

The fact that the DMD discovers underlying physics in a data-driven way without any prior knowledge of the governing equations resembles the way machine learning is used for scientific discovery. Therefore, our work has both aspects of traditional reduced-order modeling and machine learning and will be a good resource for following studies on model reduction of dynamical systems in both fields.

\begin{ack}
This work was performed under the auspices of the U.S. Department of Energy
by Lawrence Livermore National Laboratory under contract DE-AC52-07NA27344
and was supported by Laboratory Directed Research and Development funding under projects 22-SI-004 and 22-SI-006.
Y. Choi was supported for this work by the U.S. Department of Energy, Office of Science, Office of Advanced Scientific Computing Research, as part of the CHaRMNET Mathematical Multifaceted Integrated Capability Center (MMICC) program, under Award Number DE-SC0023164. IM release: LLNL-CONF-855140
\end{ack}


\bibliographystyle{ieeetr}
\bibliography{refs}

\begin{thebibliography}{10}

\bibitem{schmid2010dynamic}
P.~J. Schmid, ``Dynamic mode decomposition of numerical and experimental
  data,'' {\em Journal of fluid mechanics}, vol.~656, pp.~5--28, 2010.

\bibitem{berkooz1993proper}
G.~Berkooz, P.~Holmes, and J.~L. Lumley, ``The proper orthogonal decomposition
  in the analysis of turbulent flows,'' {\em Annual review of fluid mechanics},
  vol.~25, no.~1, pp.~539--575, 1993.

\bibitem{carlberg2018conservative}
K.~Carlberg, Y.~Choi, and S.~Sargsyan, ``Conservative model reduction for
  finite-volume models,'' {\em Journal of Computational Physics}, vol.~371,
  pp.~280--314, 2018.

\bibitem{choi2020gradient}
Y.~Choi, G.~Boncoraglio, S.~Anderson, D.~Amsallem, and C.~Farhat,
  ``Gradient-based constrained optimization using a database of linear
  reduced-order models,'' {\em Journal of Computational Physics}, vol.~423,
  p.~109787, 2020.

\bibitem{choi2019space}
Y.~Choi and K.~Carlberg, ``Space--time least-squares petrov--galerkin
  projection for nonlinear model reduction,'' {\em SIAM Journal on Scientific
  Computing}, vol.~41, no.~1, pp.~A26--A58, 2019.

\bibitem{choi2021space}
Y.~Choi, P.~Brown, W.~Arrighi, R.~Anderson, and K.~Huynh, ``Space--time reduced
  order model for large-scale linear dynamical systems with application to
  boltzmann transport problems,'' {\em Journal of Computational Physics},
  vol.~424, p.~109845, 2021.

\bibitem{hoang2021domain}
C.~Hoang, Y.~Choi, and K.~Carlberg, ``Domain-decomposition least-squares
  petrov--galerkin (dd-lspg) nonlinear model reduction,'' {\em Computer methods
  in applied mechanics and engineering}, vol.~384, p.~113997, 2021.

\bibitem{mcbane2021component}
S.~McBane and Y.~Choi, ``Component-wise reduced order model lattice-type
  structure design,'' {\em Computer methods in applied mechanics and
  engineering}, vol.~381, p.~113813, 2021.

\bibitem{choi2020sns}
Y.~Choi, D.~Coombs, and R.~Anderson, ``Sns: A solution-based nonlinear subspace
  method for time-dependent model order reduction,'' {\em SIAM Journal on
  Scientific Computing}, vol.~42, no.~2, pp.~A1116--A1146, 2020.

\bibitem{copeland2022reduced}
D.~M. Copeland, S.~W. Cheung, K.~Huynh, and Y.~Choi, ``Reduced order models for
  lagrangian hydrodynamics,'' {\em Computer Methods in Applied Mechanics and
  Engineering}, vol.~388, p.~114259, 2022.

\bibitem{kim2021efficient}
Y.~Kim, K.~Wang, and Y.~Choi, ``Efficient space--time reduced order model for
  linear dynamical systems in python using less than 120 lines of code,'' {\em
  Mathematics}, vol.~9, no.~14, p.~1690, 2021.

\bibitem{fries2022lasdi}
W.~D. Fries, X.~He, and Y.~Choi, ``Lasdi: Parametric latent space dynamics
  identification,'' {\em Computer Methods in Applied Mechanics and
  Engineering}, vol.~399, p.~115436, 2022.

\bibitem{lauzon2022s}
J.~T. Lauzon, S.~W. Cheung, Y.~Shin, Y.~Choi, D.~M. Copeland, and K.~Huynh,
  ``S-opt: A points selection algorithm for hyper-reduction in reduced order
  models,'' {\em arXiv preprint arXiv:2203.16494}, 2022.

\bibitem{cheung2023local}
S.~W. Cheung, Y.~Choi, D.~M. Copeland, and K.~Huynh, ``Local lagrangian
  reduced-order modeling for the rayleigh-taylor instability by solution
  manifold decomposition,'' {\em Journal of Computational Physics}, vol.~472,
  p.~111655, 2023.

\bibitem{mcbane2022stress}
S.~McBane, Y.~Choi, and K.~Willcox, ``Stress-constrained topology optimization
  of lattice-like structures using component-wise reduced order models,'' {\em
  Computer Methods in Applied Mechanics and Engineering}, vol.~400, p.~115525,
  2022.

\bibitem{tsai2023accelerating}
P.-H. Tsai, S.~W. Chung, D.~Ghosh, J.~Loffeld, Y.~Choi, and J.~L. Belof,
  ``Accelerating kinetic simulations of electrostatic plasmas with
  reduced-order modeling,'' 2023.

\bibitem{kutz2016dynamic}
J.~N. Kutz, S.~L. Brunton, B.~W. Brunton, and J.~L. Proctor, {\em Dynamic mode
  decomposition: data-driven modeling of complex systems}.
\newblock SIAM, 2016.

\bibitem{cheung2023datascarce}
S.~W. Cheung, Y.~Choi, H.~K. Springer, and T.~Kadeethum, ``Data-scarce
  surrogate modeling of shock-induced pore collapse process,'' 2023.

\bibitem{huhn2023parametric}
Q.~A. Huhn, M.~E. Tano, J.~C. Ragusa, and Y.~Choi, ``Parametric dynamic mode
  decomposition for reduced order modeling,'' {\em Journal of Computational
  Physics}, vol.~475, p.~111852, 2023.

\bibitem{choi2019accelerating}
Y.~Choi, G.~Oxberry, D.~White, and T.~Kirchdoerfer, ``Accelerating design
  optimization using reduced order models,'' {\em arXiv preprint
  arXiv:1909.11320}, 2019.

\bibitem{wen2023globally}
T.~Wen and M.~J. Zahr, ``A globally convergent method to accelerate large-scale
  optimization using on-the-fly model hyperreduction: application to shape
  optimization,'' {\em Journal of Computational Physics}, vol.~484, p.~112082,
  2023.

\bibitem{rocha2020adaptive}
I.~Rocha, F.~van~der Meer, and L.~J. Sluys, ``An adaptive domain-based pod/ecm
  hyper-reduced modeling framework without offline training,'' {\em Computer
  Methods in Applied Mechanics and Engineering}, vol.~358, p.~112650, 2020.

\bibitem{huang2023predictive}
C.~Huang and K.~Duraisamy, ``Predictive reduced order modeling of chaotic
  multi-scale problems using adaptively sampled projections,'' {\em arXiv
  preprint arXiv:2301.09006}, 2023.

\bibitem{fritzen2019fly}
F.~Fritzen, M.~Fern{\'a}ndez, and F.~Larsson, ``On-the-fly adaptivity for
  nonlinear twoscale simulations using artificial neural networks and reduced
  order modeling,'' {\em Frontiers in Materials}, vol.~6, p.~75, 2019.

\bibitem{rocha2021fly}
I.~Rocha, P.~Kerfriden, and F.~van Der~Meer, ``On-the-fly construction of
  surrogate constitutive models for concurrent multiscale mechanical analysis
  through probabilistic machine learning,'' {\em Journal of Computational
  Physics: X}, vol.~9, p.~100083, 2021.

\bibitem{brand2006fast}
M.~Brand, ``Fast low-rank modifications of the thin singular value
  decomposition,'' {\em Linear algebra and its applications}, vol.~415, no.~1,
  pp.~20--30, 2006.

\bibitem{matsumoto2017fly}
D.~Matsumoto and T.~Indinger, ``On-the-fly algorithm for dynamic mode
  decomposition using incremental singular value decomposition and total least
  squares,'' {\em arXiv preprint arXiv:1703.11004}, 2017.

\bibitem{zhang2019online}
H.~Zhang, C.~W. Rowley, E.~A. Deem, and L.~N. Cattafesta, ``Online dynamic mode
  decomposition for time-varying systems,'' {\em SIAM Journal on Applied
  Dynamical Systems}, vol.~18, no.~3, pp.~1586--1609, 2019.

\bibitem{mfem-web}
``{MFEM}: Modular finite element methods {[Software]}.'' \url{mfem.org}.

\bibitem{tomboulides1997numerical}
A.~Tomboulides, J.~Lee, and S.~Orszag, ``Numerical simulation of low mach
  number reactive flows,'' {\em Journal of Scientific Computing}, vol.~12,
  pp.~139--167, 1997.

\bibitem{franco2020high}
M.~Franco, J.-S. Camier, J.~Andrej, and W.~Pazner, ``High-order matrix-free
  incompressible flow solvers with gpu acceleration and low-order refined
  preconditioners,'' {\em Computers \& Fluids}, vol.~203, p.~104541, 2020.

\bibitem{williamson1989oblique}
C.~H. Williamson, ``Oblique and parallel modes of vortex shedding in the wake
  of a circular cylinder at low reynolds numbers,'' {\em Journal of Fluid
  Mechanics}, vol.~206, pp.~579--627, 1989.

\end{thebibliography}

\end{document}